\documentclass[10pt,conference,compsocconf]{IEEEtran}
\ifCLASSOPTIONcompsoc
  \usepackage[nocompress]{cite}
\else
  \usepackage{cite}
\fi
\usepackage{cleveref}

\ifCLASSINFOpdf
 \usepackage[pdftex]{graphicx}
  \graphicspath{{.}}
   \DeclareGraphicsExtensions{.pdf}
\else
\fi
\usepackage{amsmath,amssymb,bbm}
\usepackage{multirow}
\usepackage{dcolumn}
\newcolumntype{d}[0]{D{.}{.}{3.3}}
\newcolumntype{e}[0]{D{.}{.}{4.3}}

\usepackage[caption=false,font=footnotesize]{subfig}

\usepackage[shortcuts,cyremdash]{extdash}

\usepackage[nolist]{acronym}

\usepackage{paralist}
\usepackage[strings]{underscore}

\newcommand{\ndepm}{d}

\newcommand{\topom}[1]{\emph{T}_{#1}}

\newcommand{\sym}[1]{\emph{#1}}
\newcommand{\deltam}[0]{\Delta\tau}

\begin{document}
%
\title{Enhanced power grid evaluation through\\ efficient stochastic model\=/based analysis}

\author{
  \IEEEauthorblockN{Giulio Masetti}
  \IEEEauthorblockA{University of Pisa and ISTI-CNR, Pisa, Italy\\
    giulio.masetti@isti.cnr.it}
}

\IEEEspecialpapernotice{(EDCC2017 Student Forum Paper, end of second year of PhD)}

%
\begin{acronym}
\acro{GS}[G-S]{Gauss-Seidel}
\acro{JFINK-GMRES}[JF-INK-GMRES]{Jacobian-free Inexact-Newton-Krylov GMRES}
\acro{JFINK-ARNOLDI}[JF-INK-Arnoldi]{Jacobian-free Inexact-Newton-Krylov Arnoldi}
\acro{QIN}[QIN]{Quasi-Inverse-Newton}
\newacro{SAN}{Stochastic Activity Network}
\newacro{SANh}[\textbf{SAN}]{Stochastic Activity Network}
\acro{AFI}{Abstract Functional Interface}
\acro{SV}{State Variable}
\acro{NR}[\emph{NARep}]{\emph{Non\=/Anonymous Rep}}
\acro{SS}[\emph{SS}]{\emph{State\=/Sharing}}
\acro{CH}[\emph{CS}]{\emph{Channel\=/Sharing}}
\acro{RO}[\emph{DARep}]{\emph{Dependency\=/Aware Replication}}
\newacro{ICT}[ICT]{Information and Communication Tecnology}
\acro{WAN}{Wide Area Network}
\newacro{INTERNETh}[\textbf{INTERNET}]{Internet network}
\acro{INTERNET}{Internet network}
\acro{SIFS}{Short Inter-Frame Space}
\acro{STA}{Station\acroextra{ in Wi-Fi network}}
\acro{SW}{Switch}
\acroindefinite{SW}{an}{a}
\acro{TOS}{Type of Service}
\acro{TSO}{Transmission System Operator}
\acro{UDP}{User Datagram Protocol}
\acro{MF}[$MF$]{malicious failure state}
\acro{MFF}[$MFF$]{malicious failure free state}
\acro{TMF}[$T\_MF$]{permanence time in the state \acs{MF}}
\acro{TMFF}[$T\_MFF$]{permanence time in the state \acs{MFF}}
\acro{PCMF}[$P\_CMF$]{probability of choosing content malicious failure in the state \acs{MFF}}
\acro{PTMF}[$P\_TMF$]{probability of choosing timing malicious failure in the state \acs{MFF}}
\acro{MFSi}[$MFSi$]{state representing the $i$-th severity level of the malicious failure in the state \acs{MFF}}
\acro{MFSn}[$MFSn$]{state representing the $n$-th worst severity level of the malicious failure in the state \acs{MFF}}
\acro{TD}[$T\_D$]{time to execute the task when a component is in the state \acs{MFF}}
\acro{TL}[$T\_L$]{threshold for the time to execute a task when a component is in the state \acs{MFF}, used to specify when the result of the task has to be considered lost}
\acro{PMFSi}[$P\_MFSi$]{probability of reaching the $i$-th severity level from the state \acs{MFF}}
\acro{PSO}{particle swarm optimization}
\acro{TNMFS}[$TNMFS$]{time to move to the next severity level}
\acro{TPMFS}[$TPMFS$]{time to move to the previous severity level}
\acro{V}[$V$]{voltage}
\acro{Vht}[$V_i(t)$]{voltage on bus $i$ at time $t$}
\acro{VR}{Voltage Regulator}
\acro{PFP}{Power Flow Problem}
\acro{PFE}{Power Flow Equation}
\acro{SG}{Smart Grid}
\acroindefinite{SG}{an}{a}
\acroindefinite{LC}{an}{a}
\acro{delta}[$\delta$]{voltage phase angle}
\acro{deltah}[$\delta_h$]{voltage phase angle of the node $h$}
\acro{Ph}[$P_h$]{active power of the node $h$}
\acro{Qh}[$Q_h$]{reactive power of the node $h$}
\acro{Vh}[$V_h$]{voltage magnitude of the node $h$}
\acro{CSYS}{Central System}
\acro{CMCS}{Central Management and Control System}
\acro{BUS}{Bus-Bar}
\acro{BUSm}[$\mathcal{BUS}$]{Bus-Bar}
\acroplural{BUS}[BUSes]{Bus-Bars}
\acro{BG}{Bulk Generator}
\acro{DG}{Distributed Generator}
\acro{DER}{Distributed Energy Resource}
\acro{PVP}{Photovoltaic Plant}
\acro{WP}{Wind Power Plant}
\acro{DS}{Distributed Storage unit}
\acro{CB}{Capacitor Bank}
\acro{CBm}[$\mathcal{CB}$]{Capacitor Bank}
\acro{DSautoi}[$DS^{auto}_i$]{the autonomy of the storage unit on the node $i$}
\acro{DScit}[$DS^{c}_i(t)$]{The current amount of electric charge stored at time $t$ on the strorage unit on the node $i$}
\acro{DScmaxi}[$DS^{cmax}_i$]{maximum capacity of the storage unit on the node $i$}
\acro{DSi}[$DS_i$]{Distributed Storage unit (battery) on the bus $i$}
\acro{DSO}{Distribution System Operator}
\acro{DSsizei}[$DS^{size}_i$]{the size of the strorage unit on the node $i$}
\acro{E}[$V$]{voltage}
\acro{ECS}{Electrical Charging Station}
\acro{CS}{Charging Spot}
\acro{EDCA}{Enhanced Distributed Channel Access}
\acro{EI}{Electrical Infrastructure}
\acro{EO}{Evolving Objects}
\acro{EPS}{Electrical Power System}
\acro{ERP}{Extended Rate PHYs}
\acro{EV}{Electrical Vehicle}
\acro{F}{frequency}
\acro{FG}{Flexible Generator}
\acroindefinite{FG}{an}{a}
\acro{FGi}[$FG_i$]{Flexible Generator on the bus $i$}
\acro{FL}{Flexible Load}
\acro{L}{Load}
\acro{FLm}[$\mathcal{FL}$]{Flexible Load}
\acro{FLi}[$FL_i$]{Flexible Load on the bus $i$}
\acro{FPDicdots}[$FP^D_i(\cdots)$]{Flexibility Pattern for power demand}
\acro{GFit}[$G^F_i(t)$]{generation forecast at time $t$ for generator on bus $i$}
\acro{HES}{Head-end System}
\acro{HV}{High Voltage}
\acro{HV-EI}{High Voltage Electrical Infrastructure}
\acro{HVG}{Power source corresponding to the high-voltage transmission lines feeding a primary substation}
\acro{I}{Current Flow}
\acro{ICT}{Information and Communications Technologies}
\acro{Ilmax}[$I_l^{max}$]{maximum current flow that a power line $l$ can carry at time $t$ whitout being overloaded}
\acro{Il}[$I_l$]{current flow associated to line $l$}
\acro{Ilt}[$I_l(t)$]{current flow associated to line $l$ at time $t$}
\acro{IP}{Internet Protocol}
\acro{IFL}{Inflexible Load}
\acro{AN}{Access Network}
\acro{LC}{Logical Controller}
\acro{Li}[$L_i$]{Non-Flexible Load on bus $i$}
\acro{LLC}{Logical Link Control}
\acro{LV}{Low Voltage}
\newacro{LVh}[\textbf{LV}]{Low Voltage}
\acro{LV-EI}{Low Voltage Electric Infrastructure}
\acro{LV-MCS}{Low Voltage Monitoring and Control System}
\acro{LVGC}{Low Voltage Grid Control}
\acro{LVL}{Load corresponding to a secondary substation fed by the medium-voltge grid}
\acro{MAC}{Medium Access Control}
\acro{MCS}{Monitoring and Control System}
\acro{MG}{Micro-Mini Generator}
\acro{MTU}{Maximum Transmission Unit}
\acro{MV}{Medium Voltage}
\acro{MV-EI}{Medium Voltage Electric Infrastructure}
\acro{MV-MCS}{Medium Voltage Monitoring and Control System}
\acro{MVG}{Power source corresponding to a medium-voltage distribution line feeding a secondary substation}
\acro{MVGC}{Medium Voltage Grid Control}
\acro{OLTC}{On Load Tap Changer}
\acro{OLTCm}[$\mathcal{OLTC}$]{On Load Tap Changer}
\acro{OPF}{Optimal Power Flow}
\acro{P}[$P$]{active power}
\acro{PDit}[$P^D_i(t)$]{actual power demand (active power) that is met at time $t$ for load on node $i$}
\acro{PGit}[$P^G_i(t)$]{actual active power generated on node $i$ at time $t$}
\acro{Pht}[$P_h(t)$]{active power on bus $h$ at time $t$}
\acro{PHY}{physical\acroextra{ (layer)}}
\acro{PL}{Power Line}
\acro{PLCP}{Physical Layer Convergence Protocol}
\acro{PLl}[$PL_l$]{power line $l$}
\acro{Pmaxi}[$P^{max}_i$]{maximum active power that a generator can supply on bus $i$}
\acro{Q}[Q]{reactive power}
\acro{QDit}[$Q^D_i(t)$]{actual reactive power demand at time $t$ for load on node $i$}
\acro{QGit}[$Q^G_i(t)$]{actualreactive power generated on node $i$ at time $t$}
\acro{Qht}[$Q_h(t)$]{reactive power on bus $h$ at time $t$}
\acro{Qmaxi}[$Q^{max}_i$]{maximum  reactive power that a generator can supply on bus $i$}
\acro{QoS}{Quality of Service}
\acro{RES}{Renewable Energy System}
\acro{RESm}[$\mathcal{RES}$]{Renewable Energy System}
\acro{T}{Transformer}
\acro{TG}[$T_G$]{oriented graph representing the topology of EI}
\acro{thetaV}[$\theta_V$]{phase angle of voltage V}
\acro{LVLCATTACKSAN}[LV\_LC\_ATTACK\_SAN]{Template \acs{SAN} model representing the malicious failures of generic logical controller \acs{LC} at low voltage level caused by an attack to the \acs{LC}}
\acro{ANSAN}[AN\_SAN]{Template \acs{SAN} model representing a generic Access Network \acs{AN}}
\acro{LVGCATTACKSAN}[LVGC\_ATTACK\_SAN]{Template \acs{SAN} model representing the malicious failures of a generic \acs{LVGC} caused by an attack to the \acs{LVGC}}
\acro{ANATTACKSAN}[AN\_ATTACK\_SAN]{Template representing the malicious failures of a generic Access Network \acs{AN} caused by an attack to the \acs{AN}}
\acro{LVESTATESAN}[LV\_ESTATE\_SAN]{Atomic SAN model representing the update of the state of \ac{LV-EI} directly triggered by failures (outages) in \ac{LV-EI}}
\acro{LVEIINITSAN}[LV\_EI\_INIT\_SAN]{Atomic SAN model representing the initialization of the \acs{SAN} models defined in \acs{LVEIM}}
\acro{LVMCSINITSAN}[LV\_MCS\_INIT\_SAN]{Atomic SAN model representing the initialization of the models defined in \acs{LVMCSM}}
\acro{LVSPSAN}[LV\_SP\_SAN]{Template \acs{SAN} model representing the actuation of the new set points triggered by \acs{LVGC}}
\acro{LVNCCSAN}[LV\_NCC\_SAN]{Template \acs{SAN} model representing the non coordinated control actions for an area under the control of \acs{LVGC}}
\acro{LVN1M}[LV\_NODE1\_M]{Template model representing the generic node starting from a line at low voltage level}
\acro{LVN2M}[LV\_NODE2\_M]{Template model representing the generic node ending to a line at low voltage level}
\acro{LVGCSAN}[LVGC\_SAN]{Template \acs{SAN} model representing a generic \acs{LVGC}}
\acro{LVARCM}[LV\_ARC\_M]{Template model representing a generic arc at low voltage level}
\acro{LVPLSAN}[LV\_PL\_SAN]{Template \acs{SAN} model representing a generic power line \acs{PL} at low voltage level}
\acro{ARC}{Generic arc component}
\acro{NODE}{Generic node component}
\acro{MVESTATESAN}[MV\_ESTATE\_SAN]{Atomic SAN model representing the update of the state of \ac{MV-EI} directly triggered by failures (outages) in \ac{MV-EI}}
\newacro{MVESTATESANh}[\textbf{MV\_ESTATE\_SAN}]{Atomic SAN model representing the update of the state of \ac{MV-EI} directly triggered by failures (outages) in \ac{MV-EI}}
\newacro{MVPLSANh}[\textbf{MV\_PL\_SAN}]{Template \acs{SAN} model representing a generic power line \acs{PL} at medium voltage level}
\acro{MVPLSAN}[MV\_PL\_SAN]{Template \acs{SAN} model representing a generic power line \acs{PL} at medium voltage level}
\newacro{MVSWSANh}[\textbf{MV\_SW\_SAN}]{Template \acs{SAN} model representing a generic \acl{SW} at medium voltage level}
\acro{MVSWSAN}[MV\_SW\_SAN]{Template \acs{SAN} model representing a generic \acl{SW} at medium voltage level}
\newacro{MVVRSANh}[\textbf{MV\_VR\_SAN}]{Template \acs{SAN} model representing a generic \acl{VR} at medium voltage level}
\acro{MVVRSAN}[MV\_VR\_SAN]{Template \acs{SAN} model representing a generic \acl{VR} at medium voltage level}
\acro{MVBUS1SAN}[MV\_BUS1\_SAN]{Template \acs{SAN} model representing a generic \acl{BUS} at medium voltage level}
\newacro{MVBUS1SANh}[\textbf{MV\_BUS1\_SAN}]{Template \acs{SAN} model representing a generic \acl{BUS} at medium voltage level}
\newacro{MVBUS2SANh}[\textbf{MV\_BUS2\_SAN}]{Template \acs{SAN} model representing a generic \acl{BUS} at medium voltage level}
\acro{MVBUS2SAN}[MV\_BUS2\_SAN]{Template \acs{SAN} model representing a generic \acl{BUS} at medium voltage level}
\newacro{MVBGSANh}[\textbf{MV\_BG\_SAN}]{Template \acs{SAN} model representing a generic \acl{BG} at medium voltage level}
\acro{MVBGSAN}[MV\_BG\_SAN]{Template \acs{SAN} model representing a generic \acl{BG} at medium voltage level}
\newacro{MVDGSANh}[\textbf{MV\_DG\_SAN}]{Template \acs{SAN} model representing a generic \acl{DG} at medium voltage level}
\acro{MVDGSAN}[MV\_DS\_SAN]{Template \acs{SAN} model representing a generic \acl{DG} at medium voltage level}
\newacro{MVFGSANh}[\textbf{MV\_FG\_SAN}]{Template \acs{SAN} model representing a generic \acl{FG} at medium voltage level}
\acro{MVFGSAN}[MV\_FG\_SAN]{Template \acs{SAN} model representing a generic \acl{FG} at medium voltage level}
\newacro{MVDSSANh}[\textbf{MV\_DS\_SAN}]{Template \acs{SAN} model representing a generic \acs{DS} at medium voltage level}
\acro{MVDSSAN}[MV\_DS\_SAN]{Template \acs{SAN} model representing a generic \acs{DS} at medium voltage level}
\newacro{MVLSANh}[\textbf{MV\_L\_SAN}]{Template \acs{SAN} model representing a generic load at medium voltage level}
\acro{MVLSAN}[MV\_L\_SAN]{Template \acs{SAN} model representing a
  generic flexible or inflexible load at medium voltage level}
\newacro{MVFLSANh}[\textbf{MV\_FL\_SAN}]{Template \acs{SAN} model representing a generic \acl{FL} at medium voltage level}
\acro{MVFLSAN}[MV\_FL\_SAN]{Template \acs{SAN} model representing a generic \acl{FL} at medium voltage level}
\newacro{MVOLTCSANh}[\textbf{MV\_OLTC\_SAN}]{Template \acs{SAN} model representing a generic \ac{OLTC} at medium voltage level}
\acro{MVOLTCSAN}[MV\_OLTC\_SAN]{Template \acs{SAN} model representing a generic \ac{OLTC} at medium voltage level}
\newacro{MVCBSANh}[\textbf{MV\_CB\_SAN}]{Template \acs{SAN} model representing a generic \acl{CB} at medium voltage level}
\acro{MVCBSAN}[MV\_CB\_SAN]{Template \acs{SAN} model representing a generic \acl{CB} at medium voltage level}
\acro{CSYSSAN}[CSYS\_SAN]{Atomic SAN model representing the central system}
\newacro{CSYSSANh}[\textbf{CSYS\_SAN}]{Atomic SAN model representing the central system}
\acro{CMCSWANSAN}[WAN\_SAN]{\acs{SAN} model representing the \acs{WAN} associated to \acs{CMCS}}
\newacro{CMCSWANSANh}[\textbf{CMCS\_WAN\_SAN}]{\acs{SAN} model representing the Wide Area Network \acs{WAN} associated to \acs{CMCS}}
\newacro{INTERNETSANh}[\textbf{INTERNET}]{\acs{SAN} model representing the \acs{INTERNET}}
\acro{INTERNETSAN}[INTERNET]{\acs{SAN} model representing the \acs{INTERNET}}
\acro{CSYSATTACKSAN}[CSYS\_ATTACK\_SAN]{\acs{SAN} model representing the malicious failures of \acs{CSYS} caused by an attack}
\newacro{CSYSATTACKSANh}[\textbf{CSYS\_ATTACK\_SAN}]{\acs{SAN} model representing the malicious failures of \acs{CSYS} caused by an attack}
\acro{CMCSWANATTACKSAN}[CMCS\_WAN\_ATTACK\_SAN]{\acs{SAN} model representing the malicious failures of \acs{CMCSWAN} caused by an attack}
\newacro{CMCSWANATTACKSANh}[\textbf{CMCS\_WAN\_ATTACK\_SAN}]{\acs{SAN} model representing the malicious failures of \acs{CMCSWAN} caused by an attack}
\acro{INTERNETATTACKSAN}[INTERNET\_ATTACK\_SAN]{\acs{SAN} model representing the malicious failures of \acs{INTERNET} caused by an attack}
\newacro{INTERNETATTACKSANh}[\textbf{INTERNET\_ATTACK\_SAN}]{\acs{SAN} model representing the malicious failures of \acs{INTERNET} caused by an attack}
\acro{MVMCSINITSAN}[MV\_MCS\_INIT\_SAN]{Atomic SAN model representing the initialization of the models defined in \acs{MVMCSM}}
\acro{MVEIINITSAN}[MV\_EI\_INIT\_SAN]{Atomic SAN model representing the initialization of the \acs{SAN} models defined in \acs{MVEIM}}
\acro{MVSPSAN}[MV\_SP\_SAN]{Template \acs{SAN} model representing the actuation of the new set points triggered by \acs{MVGC}}
\newacro{MVSPSANh}[\textbf{MV\_SP\_SAN}]{Template \acs{SAN} model representing the actuation of the new set points triggered by \acs{MVGC}}
\newacro{MVGCSANh}[\textbf{MVGC\_SAN}]{Template \acs{SAN} model representing a generic \acs{MVGC}}
\acro{MVGCSAN}[MVGC\_SAN]{Template \acs{SAN} model representing a generic \acs{MVGC}}
\newacro{WANATTACKSANh}[\textbf{WAN\_ATTACK\_SAN}]{Template representing the malicious failures of a generic Wide Area Network \acs{WAN} caused by an attack to the \acs{WAN}}
\acro{WANATTACKSAN}[WAN\_ATTACK\_SAN]{Template representing the malicious failures of a generic Wide Area Network \acs{WAN} caused by an attack to the \acs{WAN}}
\acro{MVGCATTACKSAN}[MVGC\_ATTACK\_SAN]{Template \acs{SAN} model representing the malicious failures of a generic \acs{MVGC} caused by an attack to the \acs{MVGC}}
\newacro{MVGCATTACKSANh}[\textbf{MVGC\_ATTACK\_SAN}]{Template \acs{SAN} model representing the malicious failures of a generic \acs{MVGC} caused by an attack to the \acs{MVGC}}
\newacro{WANSANh}[\textbf{WAN\_SAN}]{Template \acs{SAN} model representing a generic Wide Area Network \acs{WAN}}
\acro{WANSAN}[WAN\_SAN]{Template \acs{SAN} model representing a generic Wide Area Network \acs{WAN}}
\acro{MVLCATTACKSAN}[MV\_LC\_ATTACK\_SAN]{Template \acs{SAN} model representing the malicious failures of generic logical controller \acs{LC} at medium voltage level caused by an attack to the \acs{LC}}
\newacro{MVLCATTACKSANh}[\textbf{MV\_LC\_ATTACK\_SAN}]{Template \acs{SAN} model representing the malicious failures of generic logical controller \acs{LC} caused by an attack to the \acs{LC}}
\acro{MVLCSAN}[MV\_LC\_SAN]{Template \acs{SAN} model representing a generic logical controller \acs{LC}}
\newacro{MVLCSANh}[\textbf{MV\_LC\_SAN}]{Template \acs{SAN} model representing a generic logical controller \acs{LC}}
\acro{MVNCCMs}[MV\_NCC\_Ms]{Composed model representing the non coordinated control actions}
\acro{MVNCCSAN}[MV\_NCC\_SAN]{Template \acs{SAN} model representing the non coordinated control actions for an area under the control of \acs{MVGC}}
\acro{MVGCM}[MVGC\_M]{Composed model representing a generic \acs{MVGC} and the malicious failures of \acs{MVGC} caused by an attack to \acs{MVGC}}
\acro{MVLCM}[MV\_LC\_M]{Template model representing a generic logical controller \acs{LC} at medium voltage level}
\acro{MVN1AN2Ms}[MV\_N1AN2\_Ms]{Composed model representing all the arcs of \acs{MV-EI} with the associated starting and ending nodes}
\acro{MVN1AN2M}[MV\_N1AN2\_M]{Template model representing a generic arc of \acs{MV-EI} with the associated starting and ending nodes}
\acro{MVN1M}[MV\_NODE1\_M]{Template model representing the generic node starting from a line at medium voltage level}
\acro{MVN2M}[MV\_NODE2\_M]{Template model representing the generic node ending to a line at medium voltage level}
\acro{MVARCM}[MV\_ARC\_M]{Template model representing a generic arc at medium voltage level}
\acro{MVGCMs}{Composed model representing all the \acs{MVGC} at medium voltage level}
\acro{MVLCMs}[MV\_LC\_Ms]{Template model representing all the logical controllers \acs{LC} at medium voltage level}
\acro{MVMCSM}[MV\_MCS\_M]{Composed model representing MV-MCS}
\acro{LVMCSM}[LV\_MCS\_M]{Composed model representing LV-MCS}
\acro{MVEIM}[MV\_EI\_M]{Composed model representing MV-EI}
\acro{LVEIM}[LV\_EI\_M]{Composed model representing LV-EI}
\acro{CMCSM}[CMCS\_M]{Composed model representing the Smart Grid at the level of CMCS}
\acro{MVM}[MV\_M]{Composed model representing the Smart Grid at the medium voltage level}
\acro{LVM}[LV\_M]{Composed model representing the Smart Grid at the low voltage level}
\acro{SM}{Smart Meter}
\acro{CEMS}{Customer Energy Management System}
\acro{CMCSWAN}{Wide Area Network associated to \acs{CMCS}}
\end{acronym}

%
%

\maketitle

\begin{abstract}
Electrical infrastructures provide services at the basis of
a number of application
sectors, several of which are critical from the perspective of human
life, environment or financials.
Following the increasing trend in electricity generation from renewable sources,
pushed by the need to meet sustainable energy goals in many countries,
more sophisticated control strategies are being adopted to regulate the
operation of the electric power system, driving electrical infrastructures
towards the so called Smart Grid scenario.
It is therefore paramount to be
assisted by technologies able to analyze the Smart Grid behavior in
critical scenarios, e.g. where cyber malfunctions or grid
disruptions occur.
In this context, stochastic model\=/based analysis are well suited
to assess dependability and quality of service related indicators, and
continuous improvements in modeling strategies and system models design
are required.
Thus, my PhD work addresses this topic by contributing to
study new Smart Grid scenarios,
concerning the advanced interplay between ICT and electrical infrastructures in
presence of cyber faults/attacks,
define a new modeling approach, based on modularity and composition,
and start to study how to improve the electrical grid dynamics representation.
In this article these studies are briefly presented and discussed.
\end{abstract}


%
\IEEEpeerreviewmaketitle

\section{Introduction and research lines description}
\label{sec:Introduction}


The complex, digital world around us requires electric power for
fundamental aspects of societal needs, business and consumer
activities.
It is therefore widely recognized that electric power
systems are among the most critical infrastructures, whose protection
is more and more a priority for many countries.
The increasing trend in electricity generation from renewable sources,
pushed by the need to meet sustainable energy goals in many countries,
poses additional challenges with the need to adopt more sophisticated
control strategies to regulate the operation of the three\=/level
electric power system: transmition, characterized by \ac{HV},
distribution, characterized by \ac{MV}, and consumer, typically
characterized by \ac{LV}.
In this panorama, studies devoted to
analyze the effectiveness of control operations and their ability to
face critical scenarios, such as in presence of failures, are
certainly well motivated.

Model-based analysis is a suitable approach to perform quantitative
estimations of a system since early stages, that is since the design
phase.
Therefore, it shows as a powerful means to support design
decision, either allowing to make the most appropriate choice among
several available alternative solutions and to facilitate tuning of
parameters when parametric solutions are employed.
Analyses devoted to assess dependability-related indicators
have already appeared in the literature.
However, the emphasis has
been mainly given to reliability and availability~\cite{ALRL04} of
\ac{ICT} infrastructure employed to guarantee power supply,
without considering explicitly
the dynamics of the underlying \ac{EI}, as for example in~\cite{ZJLS11},
or lightly introducing it.
Other studies, instead, focused mainly on the grid infrastructure,
assessing, for instance, survivability~\cite{MMDSASTMH12}
or reliability as defined in the electric sector, i.e., the ability
of the power system to deliver electricity
in the quantity and with the quality demanded by users,
as in~\cite{CF08},
neglecting the cyber control system and communications.

A more comprehensive viewpoint, targeting the interplay between
the cyber control system with the underlying controlled grid, is needed,
especially when failures occur and propagate their effects from one
level to the other.
As contribution in this direction, the SEDC
research group at ISTI\=/CNR has been working in the last years on a
stochastic modeling framework to perform quantitative estimations of
resilience-related indicators, accounting for failure events and
interdependencies between \ac{EI} and \ac{ICT}.
The outcome of these
analyses are especially helpful to understand the dynamics of relevant
phenomena and the reaction of critical components to them, so to
provide guidelines towards design improvements.
Our modeling
framework, from now on called \acs{SG} model,
is based on \ac{SAN}~\cite{SM01} formalism and the analysis
is performed via simulation so that domain specific (probability)
measures are obtained with statistical inference.
Although at a
suitable level of abstraction to cope with the inherent complexity of
the modeling effort and related solution methods, the framework
accounts for both the \ac{EI} and its \ac{ICT} distribution control
system, both at \ac{MV} and \ac{LV} level, to properly capture the
impact of dependencies among the system components.
The focus is on
the analysis of accidental faults, malicious attacks and their
propagation through existing interdependencies.
Notwithstanding the
great effort already invested in this modeling framework, the current
implementation still needs enhancements.
Especially, the ability to
address large grid topologies is at the moment rather limited.

To this purpose, advancements in the adopted solutions would be greatly
beneficial, in terms of both structural approaches supporting the
development models and analytical solvers.
This is the context where
my PhD thesis intends to provide contributions.
In particular, two major aspects of our framework
that strongly impact on performance, and so the ability to
tackle large grid infrastructures, have been identified:
\begin{itemize}
\item\label{item:IntroductionModelComposer} the model composition operator,
  at the basis of a modular
  modeling approach as adopted in the framework,
\item\label{item:IntroductionPFP} the \ac{PFP} solution strategy,
  essential for the electrical grid state estimation.
\end{itemize}
Both definition of new model composition operator and
\ac{PFP} solution strategy are
considered in my proposal, and of course
these are assumed as basic starting points for the planned research
investigations.
Another category of enhancements is new \emph{scenarios} and
measures definition and analysis.
In fact, the main research line promoted by our lab at SEDC\=/ISTI
is the development of realistic scenarios of accidental failures
or intentional attacks
and the analysis of their impact on control operations.
In nowadays distribution \acp{EI}, control
strategies need to be tested upon dynamic environmental changes and
with respect to a gamut of measures (voltage quality,
demand fulfillment, power losses, propagation of blackouts, etc.),
thus our framework has to be continuously refined to
address analysis of sophisticated grid configurations and failure models.

\begin{figure}[htbp]
\centerline{\includegraphics[scale=0.7]{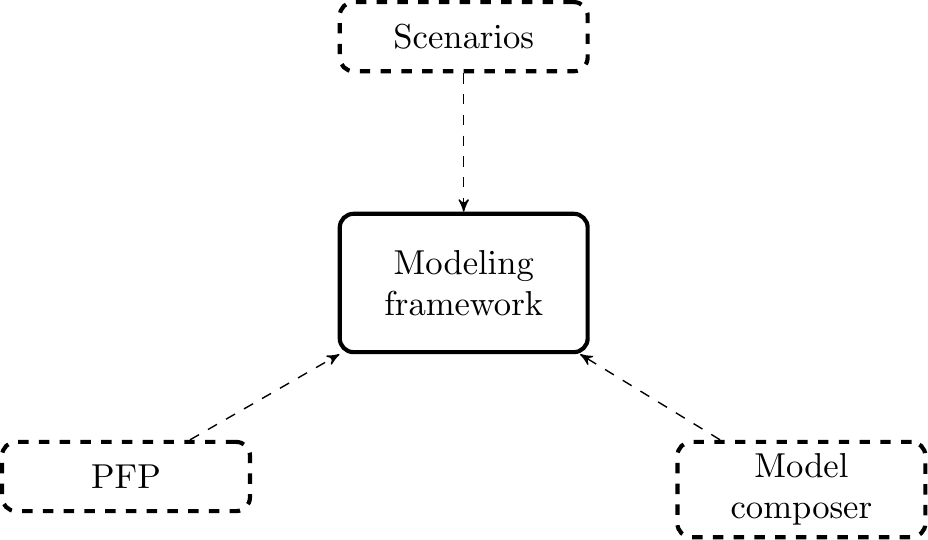}}
\caption[]{\label{fig:ResearchLines} Research lines, dashed boxes,
  with respect to
  \ac{SG} model. \ac{PFP} and Model composer
  have a direct impact on performance,
  while considering new scenarios enlarges model complexity and
  consequently can have also an impact on performance.}
\end{figure}

Summing up, my PhD work is focused on three research lines:
new scenarios definition and analysis,
new model composer strategies definitions and implementation,
and the study of \ac{PFP} solution methods, as depicted in
Figure~\ref{fig:ResearchLines}.

Structure of the paper:
\Cref{sec:Scenarios} presents interesting scenarios studied so far, describing
the effects of faults or attacks originating from \ac{EI} or
\ac{ICT} components;
\Cref{sec:ModelComposer} briefly discusses three modeling choices
we have already tested on a more general
case study with the aim to select the best from the performance point of view;
\Cref{sec:PFP} presents the \ac{PFP} and discusses investigation directions; 
in \Cref{sec:Conclusions} conclusions are drawn and future work is sketched out.


\section{\acl{SG} scenarios definition and analysis}\label{sec:Scenarios}

The focus is on the \ac{MV} level, that is composed by the
\ac{MV-EI} and the \ac{MV-MCS}.
Considered complex control policies pose our model in
the so called \ac{SG} scenario~\cite{SGIPCSWG2010}. In the following
the name \ac{SG} will refer to both \ac{EI} and \ac{MCS} together.
From a modeling point of view, the \ac{MV-EI} can be represented as
a radial or partially meshed graph, where:
\begin{itemize}
\item an arc represents a power line with the associated switch,
  \ac{OLTC} (transformer having voltage regulator) and protection
  breakers, if any;
\item each node is structured like a \ac{BUS} with the associated
  electrical equipment. Those considered in the proposed modeling
  framework are:
\begin{itemize}
\item \ac{DG}: Volatile small-scale energy generating unit, producing
  electricity from, e.g., \ac{RES} (such as wind, hydro, solar or
  photovoltaic). It can offer flexibility in the power profile,
  through power curtailment or re-dispatch.
\item \ac{IFL}: Classic load for which a loss of power is a blackout.
\item \ac{FL}: Load that offers flexibility in the power
  profile. Electrical charging stations can be considered an example
  of flexible load.
\end{itemize}
\end{itemize}

Thus, both integer and real state variables are employed
in order to capture the \ac{MV-EI} dynamics in continuous time.
In addition, the \ac{MV-EI} state is evaluated via the solution of a \ac{PFP}.
These aspects pose modeling challenges and motivates the choice of
\ac{SAN} formalism.

\begin{figure}[htp]
  \centering
  \includegraphics[scale=0.50]{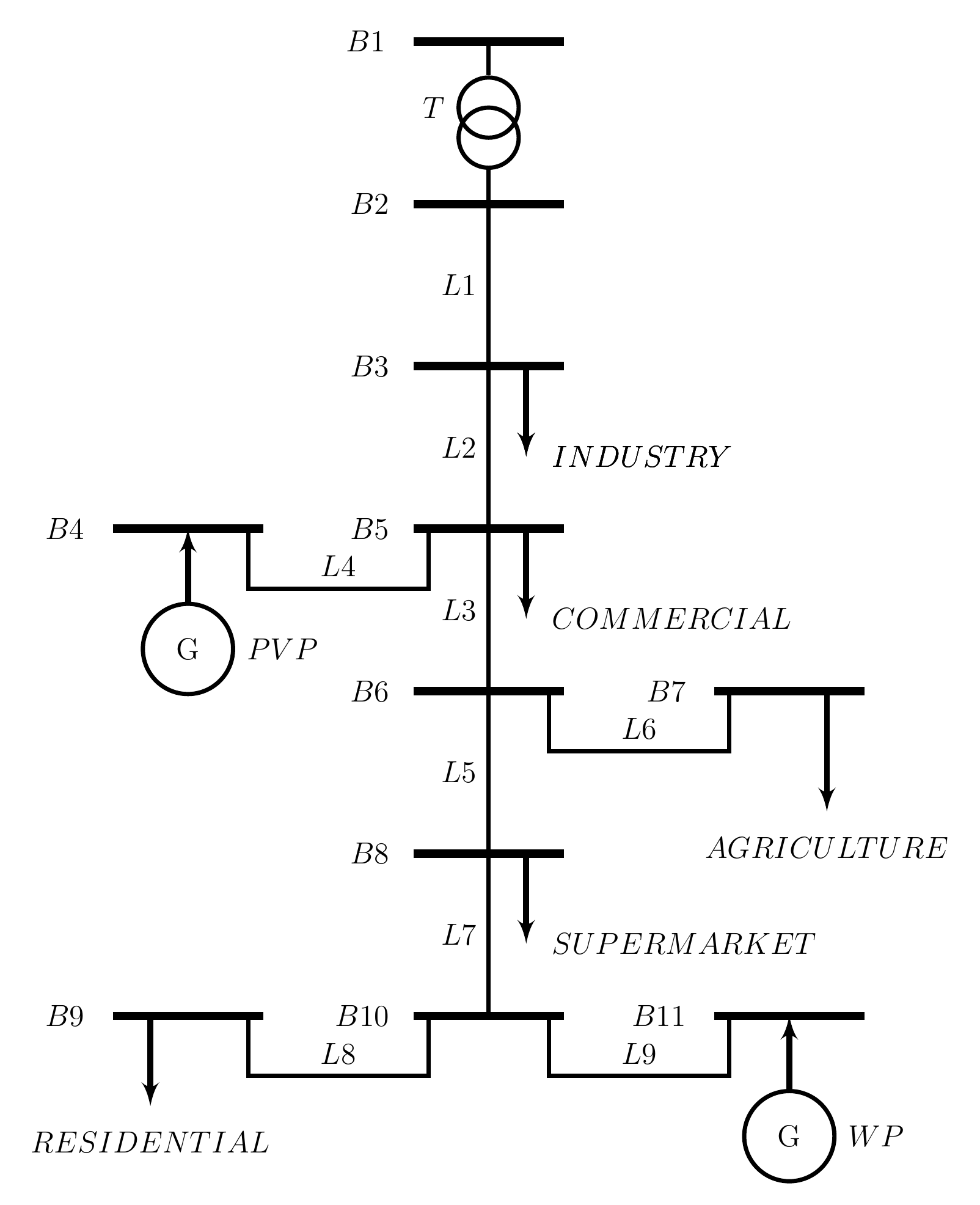}
  \caption{Diagram of a \ac{MV} smart grid from~\cite{CDGM16} (the communication
    layer is not shown).}
  \label{fig:smartc2netGrid}
\end{figure}

As an example, consider the grid shown in Figure~\ref{fig:smartc2netGrid},
taken from~\cite{CDGM16},
that is composed of $11$ \acp{BUS}, $10$ power lines,
one \ac{OLTC} between \acp{BUS} $B1$ and $B2$,
two \acp{DG} (photovoltaic at \ac{BUS} $B4$ and wind at \ac{BUS} $B11$),
and five loads, among which four are \acp{IFL} and one (\sym{INDUSTRY}
at \ac{BUS} $B3$) is \ac{FL}. 

The \ac{MV-MCS} is supposed to have a perfect knowledge of the \ac{MV-EI}
state and control actions are performed after an optimization problem is solved. 
These control actions pose modeling challenges that are addressed by
our \ac{SG} model taking advantages in particular of \ac{SAN}
gates \cite{CDGM16,CDGM16a}.

One aspect that has been emphasized in conceiving the modeling framework is the ability to account for a variety of failures, involving either the cyber control, or the grid infrastructure, or both.
At the moment, only the \emph{effect} on \ac{MV-MCS} and \ac{MV-EI} of failures are
modeled,
e.g., if the communication link between the \ac{MV-MCS} and the \ac{OLTC}
fails at a given time instant then the voltage drop at the ends
of the transformer is considered fixed from that moment on,
but details about how and why the link has failed are not modeled.
Once a failure occurs, its propagation inside the system is accounted for and the resulting impact evaluated.
The analyses progressed by first considering the presence of individual failures, and then enlarging the failure events, to also appreciate the effects of simultaneous combinations thereof.  
In this paper, only failures affecting the cyber infrastructure responsible for the distribution grid control are presented; specifically, three types of failure have been considered in~\cite{CDGM16}:
\begin{itemize}
\item timing malicious failure, 
  modeled as delayed/omitted application of
  (part of) the control actions;
\item control device failure, modeled as an incomplete application of the control actions. Specifically, the failure of control devices local to the distributed energy resources is tackled, leading to lack of control on the produced power and unavailability to perform curtailment of production to assure energy balancing;
\item \ac{OLTC} failure, potentially resulting in unsuccessful voltage control since \ac{OLTC} constitutes a major device through which voltage regulation is performed. 
\end{itemize}

We have also considered the \ac{LV} part of the grid~\cite{CDGM16a},
studying different failures, and in general the interaction between
\ac{MV} and \ac{LV},
but in the following only a \ac{MV} case study is reported to demonstrate
the potentialities of our \ac{SG} model.

The developed stochastic model-based analysis is suited to assess a variety of measures of interest to final customers, service providers and system operators. Given the interest in the voltage control functionality and its ability to promote resilient grid operation through fulfillment of voltage requirements, the following indicators have been evaluated: 
\begin{enumerate}
\item the voltage \acs{Vht} on bus $i$ measured at each time
  instant $t$ within the considered analysis period;
\item \label{ite:uov} the probability that the value of \acs{Vht} on
  bus $i$ is out of bound of
  the nominal voltage: either undervoltage $UV_i(t)$ or overvoltage $OV_i(t)$;
\item \label{ite:pmv1} the probability  $P^{\widetilde{MV1}}_i$
  that the $10$ min mean value of the supply voltage must
  be within $10 \%$ of the nominal voltage for $99\%$ of the time, evaluated over a week
  is \emph{not} met on bus $i$
  (in order to simplify the analysis the requirement has been
  evaluated over the considered analysis interval of $24$ hours);
\item \label{ite:ud} the average unsatisfied power demand $UD_i(t)$ on
  load $i$ at each instant of time $t$.
\item \label{ite:ca} the average curtailment of available power
  $CA_i(t)$ on generator $i$ at each
  instant of time $t$;
 \end{enumerate}

\begin{figure}[htp]
  \centering
  \includegraphics[scale=0.65]{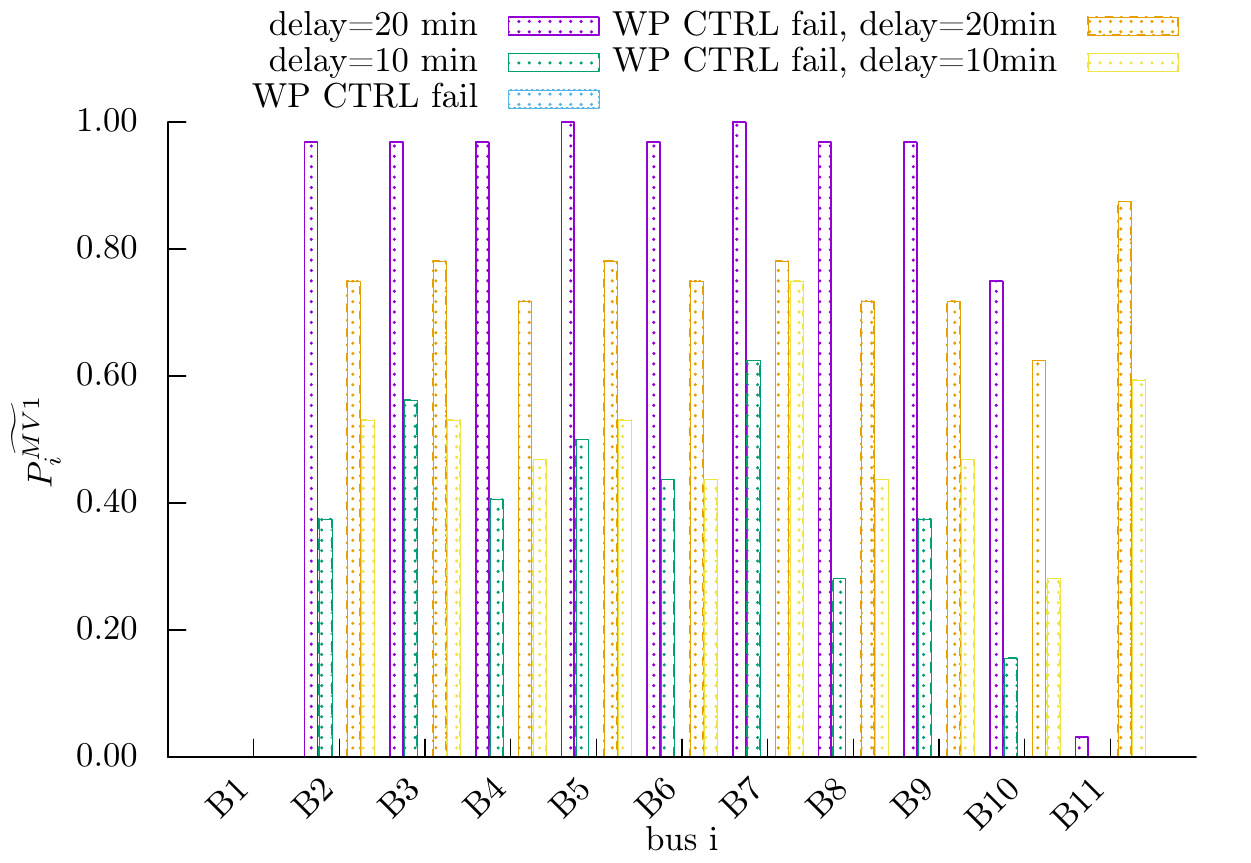}
  \caption{Probability $P^{\widetilde{MV1}}_i$ that the grid voltage requirement is not met,
    for all buses in the grid, for two
    different values of timing failure (10 min and 20 min), when 
    failures of the control device of \ac{WP} occur and when they do
    not occur.}
  \label{fig:127-measure-TVERROUT-WPCTRLdeviceFailure-tikzsa}
\end{figure}

Metrics~\ref{ite:uov}) and~\ref{ite:pmv1}) are
representative of the degree of reliability of the smart
grid in delivering its service, while metrics~\ref{ite:ud}) and~~\ref{ite:ca}) express
the effectiveness of the analyzed voltage control functionality
in satisfying customers expectations.
As an example, in Figure~\ref{fig:127-measure-TVERROUT-WPCTRLdeviceFailure-tikzsa}
measure~\ref{ite:pmv1}), i.e., the probability $P^{\widetilde{MV1}}_i$
that the voltage requirement is not met, is depicted for every \ac{BUS} $i$ of the grid illustrated in Figure~\ref{fig:smartc2netGrid}, comparing
the impact of timing failure with respect to failures of the control device of \ac{WP}. 

In order to compute measures~\ref{ite:uov}) and~\ref{ite:pmv1})
with a reasonably small confidence interval (e.g., $10^{-5}$) exercising grids of the
size of Figure~\ref{fig:smartc2netGrid} or with $48$ \acp{BUS}, as in~\cite{CDGM16a},
several hours of computation are needed on a Intel(R) Core(TM) i7-5960X with fixed 3.50 GHz CPU,
20M cache and 32GB RAM, an up to date GNU/Linux Operating System and using the M\"{o}bius
Modeling Framework~\cite{DCCDDDSW02}.
Being interested in addressing electrical grids with hundreds or thousands \acp{BUS},
such as in the IEEE118, IEEE300 testbed~\cite{IEEE118,IEEE300}
and the Illinois Center for a Smarter Electric Grid's Texas synthetic grid~\cite{Texas2000}, directions for improvements are presented in \Cref{sec:ModelComposer,sec:PFP}.


\section{New modeling strategies definition}\label{sec:ModelComposer}

Abstracting away from the \ac{SG} scenario,
the logical structure of the considered systems comprises:
\begin{itemize}
\item A large number of cyber\=/physical components,
  weakly interconnected with each other according to
  physical and cyber topologies.
\item One or more generic components.
  Each generic component groups all the specific components
  having common characteristics, i.e., homogeneous system components,
  which, although different, 
  share the same behaviour, structure and parameters.
  This means that a template model built for the generic component
  is adequate to represent the set of its specific components.
\item A central \ac{MCS} capable to communicate with each specific
  component.
\end{itemize} 

As examples of weakly interconnected electrical components,
in the IEEE118, IEEE300 testbed~\cite{IEEE118,IEEE300}
and the Illinois Center for a Smarter Electric Grid's Texas synthetic grid~\cite{Texas2000},
the interconnection degrees are numbers between $2$ and $3$ on average,
with maximum value of $16$ for the configuration with $2000$ nodes.
Electrical nodes are representable as instances of a generic component, called \ac{BUS};
different electrical components, e.g., \acp{DG} and \acp{OLTC}, can be attached to each \ac{BUS},
thus electrical nodes are identified by their position in the electrical grid and the
list of components attached on them.
In the rest of the paper, as an example of communication topology, we will consider
the \ac{MV-MCS} connected directly to all the \ac{MV} electrical components.

In order to describe how the system logical structure is translated in our \ac{SG} model,
why the model can not scale at increasing the number of electrical nodes and my proposal
for a new strategy to overcome the problem, some additional information concerning the modeling
formalism and composition operators are needed.
As already mentioned in \Cref{sec:Introduction}, we opt for
the \ac{SAN} formalism~\cite{SM01}, a stochastic extension of Petri nets
based on four primitives: places,
activities (transitions), input gates, and output gates.
Primitive data types of the programming language
C++, like short, float, double, including structures and arrays, are
represented by special places, called ``extended places''.
Input gates define both the enabling condition of an activity and the
marking changes occurring when the activity completes.
The output gates define the marking changes occurring when the
activity completes, but they are randomly chosen at completion of the activity 
from a probability distribution function, defined by ``cases''
associated to the activity.
The modeler defines input and output gates writing chunks of C++ code,
thus having a great expression power.
Composed models are obtained through two compositional operators, based 
on the sharing of places~\cite{SM91}:
\begin{itemize}
\item \emph{Join}, composes, i.e., brings together two or
  more (composed or atomic) submodels.
  The expression
  \begin{equation*}
    M = \mathcal{J}\Big(\{p_{1},\dots,p_{m}\};\, SM_{1},\dots,SM_{n}\Big)
  \end{equation*}
  means that a new model $M$ is created by the juxtaposition
  of submodels $SM_{1}$, \dots, $SM_{n}$ and if the place $p_{j}\in\{p_{1},\dots,p_{m}\}$ appears in
  more then one submodel then $p_{j}$ will appear only once in $M$,
  maintaining all the arcs that connect $p_{j}$ to activities and gates.  
\item \emph{Rep}, automatically constructs identical copies
  (replicas) of a (composed or atomic) submodel.
  The expression
  \begin{equation*}
    M = \mathcal{R}_{n}\Big(\{p_{1},\dots,p_{m}\};\, SM\Big)
  \end{equation*}
  means that the submodel $SM$ is copied $n$ times and the places
  $\{p_{1},\dots,p_{m}\}$ are all shared among all the replicas.
\end{itemize}

In~\cite{CDGM16b}, issues in modeling a large population of similar
and weakly interconnected
components were introduced and \emph{NARep} served as starting point
for the following discussion.
We have identified three different modeling strategies that match the
system logical structure.
Starting from a model of the generic cyber\=/physical component,
these strategies guide the automatic definition of $n$ specific components:

\begin{itemize}
\item \ac{SS}: the generic component model \sym{GENCOMP} comprises
  an indexing mechanism, the index\=/dependent behaviour model
  and the set $\{s_{1},\dots,s_{n}\}$ of places, where $s_{j}$ describes
  the portion of component $j$ state that is relevant for some other
  component.
  \sym{GENCOMP} is replicated $n$ times and
  $\{s_{1},\dots,s_{n}\}$ is globally accessible. Formally:
  \begin{equation*}
    \mathcal{R}_{n}\Big(\{s_{1},\dots,s_{n}\};\, GENCOMP\Big)
  \end{equation*}
  This strategy, already presented in~\cite{F12}, is momentary implemented in our
  \ac{SG} model~\cite{CDGM16}.
  It is a general solution, but its efficiency is limited by the fact that
  it assumes a complete graph of interactions among the replicated components.
  This assumption does not match with the great majority of real\=/world
  systems,
  typically composed by many loosely interconnected components according to
  regular dependency topologies (tree, mesh, cycle, etc).
\item \ac{CH}: the generic component model \sym{GENCOMP} comprises
  an indexing mechanism, the index\=/dependent behaviour model, 
  a communication channel \sym{ch} and the submodel \sym{CHMAN} that manage the
  channel. Formally:
  \begin{equation*}
    \mathcal{R}_{n}\Big(\{ch\};\, \mathcal{J}\Big(\{ch\};\, GENCOMP,CHMAN\Big)\Big)
  \end{equation*}
  \sym{GENCOMP} is replicated $n$ times but only \sym{ch}, a light extended place,
  is shared among all the replicas. \sym{CHMAN} regulates the channel usage and maintains
  in sync, inside each replica, a copy of portion of other components' state.
  Synchronizations take place by means of instantaneous actions, thus dependability
  measures are not impacted by \sym{CHMAN}, but the price to pay is the increase of the
  events number.
  Details about this strategy and comparisons with
  \ac{SS} will appear in EPEW2017 workshop proceedings~\cite{MCDG17a}.
\item \ac{RO}: the generic system component is modeled by means of the template
  model \sym{TEMPLATE} and the interdependency topology $\topom{n}$.
  Starting from \sym{TEMPLATE} and $\topom{n}$, $n$ new models $COMP_{1},\dots,COMP_{n}$
  are automatically created, where $COMP_{i}$ contains $s_{j}$,
  a portion of component $j$ state, only if component $i$ depends
  on component $j$. All the component models are joined together, but each $s_{j}$ is shared
  only among those models that need it. Formally:
  \begin{equation*}
    \mathcal{J}\Big(\{s_{1},\dots,s_{n}\};\, COMP_{1},\dots,COMP_{n}\Big)
  \end{equation*}
  A new composition operator that is capable
  to produce $COMP_{1},\dots,COMP_{n}$ starting from \sym{TEMPLATE} and
  $\topom{n}$ have been defined; its implementation, in conjunction with the
  M\"{o}bius framework, is based on XQuery~\cite{B04}.
Details about this operator will appear in ISSRE2017 conference proceedings
~\cite{MCDG17b}.
\end{itemize}

All three approaches have been tested on a case study that
is effective in demonstrating the features of the mechanisms,
and can be considered as a basis to be easily extended and adapted to represent
a great variety of real contexts, far beyond the \ac{SG} scenario.
We have considered $n$ working stations dedicated to perform
the same task in parallel.
At every time instant, each station can be either \emph{working} or
\emph{failed}, and the change of status takes place after an exponentially
distributed random time.
The failure of a station implies a reconfiguration of the workload
assigned to the other stations, to continue accomplishing the tasks
of the failed station.
Just before failing, a station redirects its
tasks to one or more other stations it is connected with,
i.e. neighbouring stations according to the dependency topology.
The stations that receive new tasks increase
their workload, implying also a change of their failure rate.
Thus, the model is a pure death process~\cite{T02}
with monotone load sharing~\cite{A08}.

We have modeled the case study following all the three strategies
(\ac{SS}, \ac{CH} and \ac{RO}) studying in particular how to transform
the model that implements \ac{SS} into models that implement \ac{CH} and
\ac{RO} in order to facilitate the change of strategy inside our \ac{SG} model.
Performance comparisons have confirmed the expected improvements.
Results of a complete analysis will appear in~\cite{MCDG17b},
and here, to illustrate the improvements, only
time measures about
\ac{SS} and \ac{RO}, obtained with the terminating simulator of
the M\"{o}bius tool~\cite{DCCDDDSW02}, are briefly discussed.
In particular, consider $\deltam(k)$ the
difference between the total amount of \sym{CPU} time, in seconds,
used by one execution of the M{\"o}bius simulator that runs $k$ batches
and the amount of \sym{CPU} time, in seconds,
used by one execution of the M{\"o}bius simulator to initialize the data
structures of the simulator.

\begin{table}[htbp]
\caption{$\deltam(1000)$ in seconds for the \ac{SS} approach.}
\label{tab:KFixedAbsoluteSS}
\centerline{
\begin{tabular}{|c|d|d|d|d|}
\hline
 & \multicolumn{1}{c|}{$\ndepm=1$} & \multicolumn{1}{c|}{$\ndepm=9$} & \multicolumn{1}{c|}{$\ndepm=99$} & \multicolumn{1}{c|}{$\ndepm=500$} \\
\hline
\multirow{1}{*}{$n=10^{1}$} & 0.087 & 0.102 &  & \\
\hline
\multirow{1}{*}{$n=10^{2}$} & 9.203 & 9.197 & 9.357 & \\
\hline
\multirow{1}{*}{$n=10^{3}$} & 1613.246 & 1723.666 & 1732.983 & 1754.996 \\
\hline
\end{tabular}
}
\end{table}

\begin{table}[htbp]
\caption{$\deltam(1000)$ in seconds for the \ac{RO} approach.}
\label{tab:KFixedAbsoluteRO}
\centerline{
\begin{tabular}{|c|d|d|d|d|}
\hline
 & \multicolumn{1}{c|}{$\ndepm=1$} & \multicolumn{1}{c|}{$\ndepm=9$} & \multicolumn{1}{c|}{$\ndepm=99$} & \multicolumn{1}{c|}{$\ndepm=500$} \\
\hline
\multirow{1}{*}{$n=10^{1}$} & 0.0015 & 0.0022 &  & \\
\hline
\multirow{1}{*}{$n=10^{2}$} & 0.774 & 0.817 & 0.972 & \\
\hline
\multirow{1}{*}{$n=10^{3}$} & 104.939 & 109.797 & 131.580 & 167.160 \\
\hline
\end{tabular}
}
\end{table}

\Cref{tab:KFixedAbsoluteSS,tab:KFixedAbsoluteRO}
depict $\deltam^{\ac{SS}}(k)$ and $\deltam^{\ac{RO}}(k)$
respectively, where $k=1000$ simulation batches
are considered for a variable number $n$ of system components, each
being dependent on a variable number $\ndepm$ of other components.

Although the values shown by \ac{RO} are very small and significantly lower than the corresponding ones of \ac{SS}, it can be observed the different trend of the two approaches with respect to $\ndepm$. In fact, while the impact of $\ndepm$ on $\deltam(1000)$ is very small in the table relative to \ac{SS}, in the case of \ac{RO} the value of $\deltam(1000)$ for  $\ndepm=500$ is about $1.6$ the value for $\ndepm=1$. This is not surprising, since \ac{SS} always works under the implicit assumption of maximum interconnection among component replicas, so its sensitivity to variation of $\ndepm$ is minimal, while \ac{RO} is influenced by $\ndepm$, given the applied principle of considering only real replicas interdependencies. With respect to increasing values of $n$, as expected the results obtained for $\deltam(1000)$ increase for both approaches. However, \ac{RO} can be about one order of magnitude faster than \ac{SS} when $n=1000$ and $\ndepm$ up to $9$.


\section{\acl{PFP} solution improvements}
\label{sec:PFP}

In our \ac{SG} model, the \ac{EI} state is determined~\cite{R.Idema2014}
from the knowledge, for all \ac{BUS} $i$, of:

\begin{itemize}
\item injected power phasor $S^{g}_{i}$,
\item demanded power phasor $S^{d}_{i}$,
\item lines characteristics (admittance bus matrix $Y_{bus}$),
\item relationship between power $S^{bus}_{i}$
and voltage $V_{i}$ given by
\begin{equation*}
S^{bus}_{i} = V_{i}\sum_{k}^{n} \big(Y^{*}_{bus}\big)_{ik}V^{*}_{k}\text{,}
\end{equation*}
\item power balance equations
\begin{equation*}
G_{i}(V_{1},\dots,V_{n}) = S^{bus}_{i} + S^{d}_{i} - S^{g}_{i} = 0\text{.} 
\end{equation*}
\end{itemize}

The \ac{PFP} consists in extracting relevant information from the equations $G(V)=0$. 
Notice that $S^{g}_{i}$ is the sum of powers produced by \acp{DG} attached at \ac{BUS} $i$ and
$S^{d}_{i}$ is the sum of powers consumed by loads attached at \ac{BUS} $i$.
Thus, during the simulation of the stochastic process described by our \ac{SG} model,
each new event, e.g. failure of a \ac{DG}, can induce different values
of power generated/requested and then a new \ac{PFP} has to be solved.
The most common strategy to solve the \ac{PFP} consists in breaking up the
complex set of non\=/linear equations $G(V)=0$ in real and imaginary parts, and in considering the polar decomposition of
voltages in order to
obtain the real set $F=0$ of $2n$ non\=/linear equations.
The Newton\=/Raphson method~\cite{R.Idema2014} is adopted to solve $F=0$
but its standard formulation is inefficient and
constitutes a relevant bottleneck during the \ac{SG} model simulation.

To mitigate the impact of this computation,
the Inexact\=/Newton\=/Krylov GMRES method~\cite{Idema2012}
have been proposed and implemented
so that a gain in scalability with respect to the
number $n$ of \acp{BUS} is expected.
The INK GMRES is impacted by the equations ordering, as preliminary discussed in~\cite{MCDG16},
and particular choices of orderings can result in even better performance.
Another strategy, originating from~\cite{Trias2012}, to solve the \ac{PFP} requires
the introduction of a
new complex parameter $s$ and the analysis of the set of \emph{functional} equations $\tilde{G}(V(s))=0$.
This strategy, still to be detailed, is a promising alternative to the Newton\=/Raphson method
in the \ac{SG} scenario because, with adequate adjustments, can handle many similar \acp{PFP}
with a single computation, with benefits in therms of computational cost over the entire model simulation.


\section{Conclusions}\label{sec:Conclusions}

Moving from considerations on the need to promote efficient dependability
and performability model\=/based analysis to properly address the increasing size
of modern and future \acp{SG}, this paper presented
three research lines I am addressing in the context of my PhD studies.
New \ac{SG} scenarios definition with high interaction between \ac{ICT} and \ac{EI},
and increasing level of details, is the main topic of my research.
It presents modeling challenges from the performance point of view and then
addresses new modeling strategies and new \ac{EI} state representation via
advanced \ac{PFP} solution strategies.
The main idea of new modeling solutions, \ac{CH} and \ac{RO}, is to exploit the actually existing dependencies among components of the system under analysis, instead of relying on the pessimistic situation of point\=/to\=/point connections as assumed by the already existing \ac{SS} approach.
The main idea of \ac{PFP} solution improvements is to re\-/think the Newton\=/Raphson method and to study the applicability of its alternatives.

Extensions of the presented studies are foreseen in several directions, some of which
are:
\begin{itemize}
\item new scenarios definition, in particular, introducing the interaction
  between \ac{HV} and \ac{MV} by means of a limited power interchange between them
  and new measures about the quality of service perceived not only by the electrical
  service provider but also by consumers,
\item tackle source of failure, not addressed at the moment,
\item re\=/design our \ac{SG} model to include the \ac{RO} approach and test it
  on large \acp{SG},
\item study different orderings of the equations arising from the \ac{PFP} to
  increase the model simulation performance,
\item study the feasibility of new \ac{PFP} solution techniques, such as elaborations
  on $\tilde{G}=0$, application in the \ac{SG} scenario,
\item apply the \ac{RO} approach in modeling system outside the \ac{SG} scenario,
\item a long term objective would be to define and making native in the adopted evaluation tool a new non\=/anonymous replica operator, based on the principle of \ac{RO}.
\end{itemize}

\section*{Acknowledgment}

I would like to thank Felicita Di Giandomenico, my advisor,
and Silvano Chiaradonna for providing insights and expertise
that are guiding my research.

\bibliographystyle{IEEEtran}
\bibliography{EDCC2017StudentForum}
%

\end{document}